\documentclass[10pt,conference,a4paper]{IEEEtran}

\usepackage{hyperref}
\usepackage{color}
\usepackage{amssymb}
\usepackage{rotating}
\usepackage[T1]{fontenc}
\usepackage{lscape}
\usepackage{dblfloatfix}
\usepackage{xcolor}
\usepackage{subcaption}
\usepackage[ruled,vlined]{algorithm2e}

\ifCLASSINFOpdf
\else
 
\fi

\begin{document}
%
\title{Appliance identification using a histogram post-processing of 2D local binary patterns for smart grid applications}


\author{\IEEEauthorblockN{Yassine Himeur, Abdullah Alsalemi, Faycal Bensaali}
\IEEEauthorblockA{\textit{Department of Electrical Engineering,} \\
\textit{Qatar University, Doha, Qatar}\\
Email: yassine.himeur@qu.edu.qa, a.alsalemi@qu.edu.qa, \\
 f.bensaali@qu.edu.qa}

\and
\IEEEauthorblockN{Abbes Amira}
\IEEEauthorblockA{\textit{Institute of Artificial Intelligence,} \\
\textit{De Montfort University, Leicester, United Kingdom}\\
Email: abbes.amira@dmu.ac.uk}
}


\maketitle

\begin{abstract}
Identifying domestic appliances in the smart grid leads to a better power usage management and further helps in detecting appliance-level abnormalities. An efficient identification can be achieved only if a robust feature extraction scheme is developed with a high ability to discriminate between different appliances on the smart grid. Accordingly, we propose in this paper a novel method to extract electrical power signatures after transforming the power signal to 2D space, which has more encoding possibilities. Following, an improved local binary patterns (LBP) is proposed that relies on improving the discriminative ability of conventional LBP using a post-processing stage. A binarized eigenvalue map (BEVM) is extracted from the 2D power matrix and then used to post-process the generated LBP representation. Next, two histograms are constructed, namely up and down histograms, and are then concatenated to form the global histogram. A comprehensive performance evaluation is performed on two different datasets, namely the GREEND and WITHED, in which power data were collected at 1 Hz and 44000 Hz sampling rates, respectively. The obtained results revealed the superiority of the proposed LBP-BEVM based system in terms of the identification performance versus other 2D descriptors and existing identification frameworks.

\end{abstract}
\begin{IEEEkeywords}
Appliance identification, 2D power representation, local binary patterns, binarized eigenvalue map, classification.
\end{IEEEkeywords}

\IEEEpeerreviewmaketitle

\section{Introduction}

Excessive energy consumption in domestic buildings became a significant issue in recent years and strict measures should be set to reduce it. Specifically, residential households can consume up to 40\% of the overall worldwide energy consumption \cite{Alsalemi8959214}. This consumption will likely be increased by 2030 due to the improved living conditions for a large number of people around the world, and hence various new electrical appliances are used in each household \cite{Devlin2019}. From another side, this increase in energy consumption is offset by a decrease of natural resources, including oil, gas and coal. Consequently, many governments are greatly concerned with establishing new energy saving strategies, especially based on the use of smart grids that are related to information and communication technologies (ICT), such as Internet of things (IoT), artificial intelligence, etc \cite{Alsalemi9112672}.

Among the conventional techniques to measure energy usage of electrical devices is the intrusive load monitoring (ILM). The latter aims at implementing individual smart-meters or sensors to collect power usage data from each appliance separately \cite{Himeur2020icict,HIMEUR2020115872}. However, this technique is very costly since it requires an enormous overhead as a result of conducting various tasks such as installing and maintaining different sub-meters, communicating between them, storing collected data and checking data validity \cite{Li2019SG}. 

Unlike ILM schemes that monitor power usage with several individual sensors, non-intrusive load monitoring (NILM) is developed to infer the device-wise consumption from an aggregated consumption that is gathered from the main supply. This results in a reduced even a null cost for installation and maintenance \cite{Welikala2019SG}. Moving forward, NILM systems play an essential role in helping end-users reducing their energy consumption through providing them with real-time consumption footprints, improving their energy usage habits, and further detecting faulty devices \cite{Liu2019SG}.

In order to develop an efficient NILM framework, feature extraction is one of the most critical tasks. Steady-state detection techniques have been used to extract characteristics in NILM system, however, this kind of approaches has the disadvantage of not being robust enough to identify  appliances with similar power usage. This usually leads to a miss-classification and poor identification performance \cite{Himeur2020iscas,Himeur2020AE}. In this context, designing a feature extraction scheme that can efficiently classify appliances having similar power usage is a current challenge.

This paper proposes an original feature extraction approach for appliance identification over the smart grid based on capturing prominent power consumption characteristics. Specifically, power signals are firstly transformed into 2D space and hence offering more possibilities to extract features using square kernels. Each sample will be surrounded by at least 8 close neighbors and different description methods can be applied, such as binary, phase and texture descriptions. In contrast to using 1D descriptors, in which only 2 close neighbors exist and a limited number of possibilities are available to extract the characteristics.
To the best of authors' knowledge, this is the first work that treats power signals in 2D space to extract representative signatures. After that, an improved local binary patterns (LBP) descriptor is applied that is mainly based on extracting the LBP representation using a conventional LBP descriptor, and in parallel, generating a preliminary information from the power matrix obtained in 2D space using a binarized eigenvalue map (BEVM). Following, this information is injected into a post-processing block that helps in extracting two new histograms from the LBP representation. Afterward, the resulting histograms are concatenated to form the global histogram. Finally, a set of experimental tests is then conducted to evaluate the proposed LBP-BEVM descriptor under two real power consumption datasets.        

The remaining sections are organized as follows. Section \ref{sec2} describes related works. Section \ref{sec3} explains the proposed methodology based on LBP-BEVM. Empirical results are presented in Section \ref{sec4}. Finally, Section \ref{sec5} concludes the paper.  

\section{Related works} \label{sec2}
Smart grid that aims at enabling a better energy managing and saving, requires also powerful tools to identify appliances and extract their individual consumption footprints. However, in order to achieve a high identification accuracy, developing  efficient feature extraction schemes providing a better discriminative ability is a primary concern. Several characteristic extraction descriptors have been investigated in the state-of-the-art.

In \cite{Ghosh8665386}, harmonic impedance features are determined on the load-side for various electrical device classes implemented in a specific house. The second step focuses on implementing a fuzzy rule-based technique to identify the individual appliances at the consumer-end. 
In \cite{Kulkarni6980072},  Kulkarni et al. collect electromagnetic field characteristics generated by various domestic devices in order to construct a distinct fingerprint for every electrical appliance. Afterward, a decision tree classifier is used to automatically manage the identification task. 
In \cite{Park8031399}, features collected from real power values of electrical devices are used along with a cogent confabulation neural network (CCNN) to identify each appliance. 
In \cite{Du6982196}, pertinent load characteristics are extracted using a finite-state-machine representation, in which the root mean square and staying time parameters are deployed to represent appliance states and events before feeding them to a classification algorithm.  

In \cite{Xiao8873557}, a statistical model is introduced to develop fingerprints of load events and hence supporting performances of anti-interference of voltage/current variations.
In \cite{Le9042316}, the authors use the magnitude and phase of the appliance current signal and the fast Fourier transform (FFT) to collect steady-state characteristics. The latter is then fed to a machine learning classifier for the identification purpose. 
In \cite{Himeur2020AE}, a multi-scale wavelet packet tree (MSWPT) is proposed to construct relevant fingerprints of each electrical appliance. Following, a set of different learning classifiers is deployed to classify the appliances.

Most of the appliance identification systems are implemented in 1D space, they are in some way lacking of robustness and are usually trained using complex deep learning models. However, in this paper, we deploy an improved 2D descriptor to extract pertinent appliance features from power signals. This idea is motivated by the fact that transforming power signals to 2D space opens more possibilities of encoding power observations and different kinds of binary descriptions can be used as well. Moreover, this descriptor has a low computation cost and can be trained with conventional machine learning classifiers without the need to use of complex deep learning models, which are usually computationally intensive and hard to implement on low-cost computing platforms. Furthermore, extracting 2D features of power signals using the proposed descriptor results in a higher identification performance in comparison with using 1D patterns or even other conventional 2D descriptors. The former can not only improve the correlation/discrimination ability of the classifier model, but it acts also as a dimensionality reduction scheme, where it encodes effectively the class-specific features via removing the unnecessary information.

On the other side, the improved 2D descriptor is tested on two different datasets, collected at completely different scenarios. The first one is collected at low frequency resolution of 1 Hz, in which daily appliance consumption signatures are collected for a period of more than three months from the same household. While in the second dataset, appliance fingerprints are collected from different appliance categories and each appliance category includes various appliances from distinct manufacturers. Demonstrating the high performance of our improved 2D descriptor proves its applicability in real application scenarios. Thus, our algorithm can identify any appliance even from a different manufacturers if it has already trained using a simple yet effective EBT classifier on the same appliance class without the need to train it again if it is deployed on real-world applications. As it is also possible that the end-user uses its own data to train it before running an online identification. In addition, this demonstrates also that in our case there is no need to conduct transfer learning when our algorithm is applied on real-world applications as it is the case when deep learning models are used \cite{Liu8580416}.

In addition, this kind of 2D descriptors which uses a simple yet efficient binary description technique could be implemented at a very low complexity using low-cost platforms (e.g. Raspberry Pi 3 or Jetson Nano), as described in \cite{RPI3book}. Explicitly, conventional LBP has been successfully applied for real-time applications using low-cost platforms for face recognition and video surveillance applications \cite{RPI3book}. On the other hand, power signals are less complex than images or videos, and hence implementing the improved 2D descriptor with the EBT classifier will have a faster time execution than other recent solutions, especially those based on deep learning, which can ensure developing real-time applications.

\section{Proposed system} \label{sec3}
\begin{figure}[t!]
\begin{center}
\includegraphics[width=9.2cm, height=3.6cm]{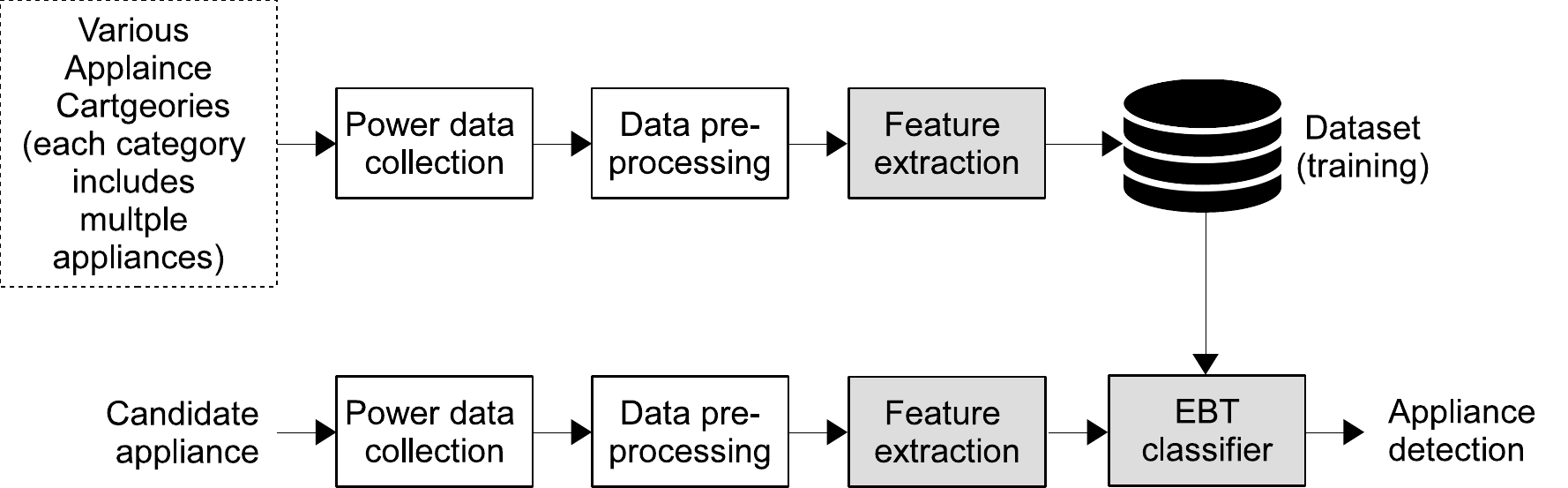}
\end{center}
\caption{Block diagram of the proposed appliance identification system.}
\label{AppIden}
\end{figure}

\begin{figure*}[t!]
\begin{center}
\includegraphics[width=17.7cm, height=3.7cm]{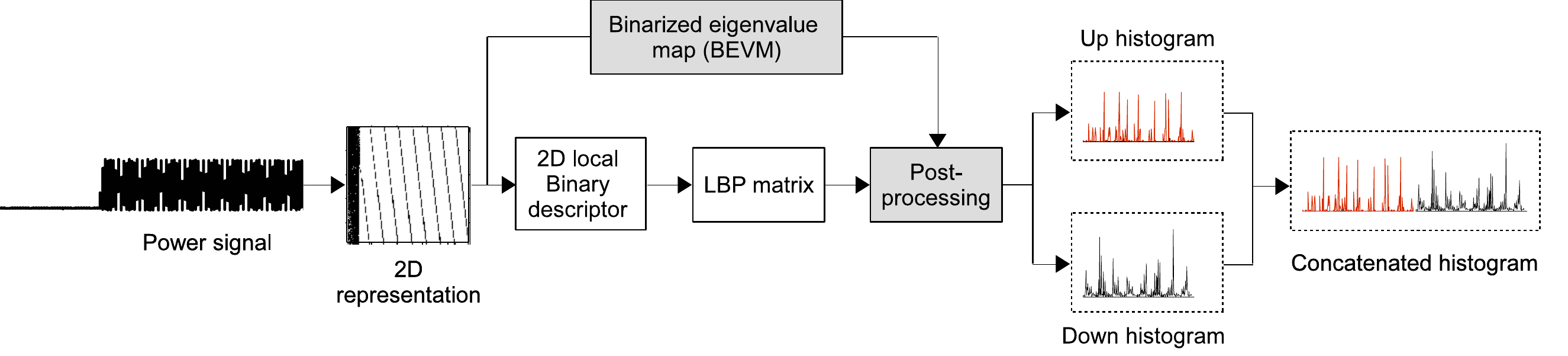}
\end{center}
\caption{Flowchart of the proposed feature extraction descriptor.}
\label{FeaExtraction}
\end{figure*}

To automatically identify electrical devices on a smart grid, the proposed methodology relies on two important parts; i.e. an improved feature extraction method based on a 2D representation of power signals and a powerful ensemble bagging tree (EBT) classier that can efficiently discriminate between the appliance features. Fig. \ref{AppIden} illustrates the block diagram of the proposed methodology.
LBP is simple descriptor that has proved its efficiency in other fields, such as face recognition \cite{Ahonen2006}, however, to use it for appliance identification some improvements are required. The flowchart of the proposed feature extraction is depicted in Fig. \ref{FeaExtraction}. As illustrated, after converting the power signal into 2D space, it is processed by LBP descriptor \cite{Ahonen2006} to extract the LBP representation. In parallel, another segment of information is generated called BEVM, which is served to post-process the generated LBP matrix in order to construct two new robust histograms. 

\subsection{Extracting the binarized eigenvalue map (BEVM)}
\begin{figure}[t!]
\begin{center}
\includegraphics[width=6cm, height=2.2cm]{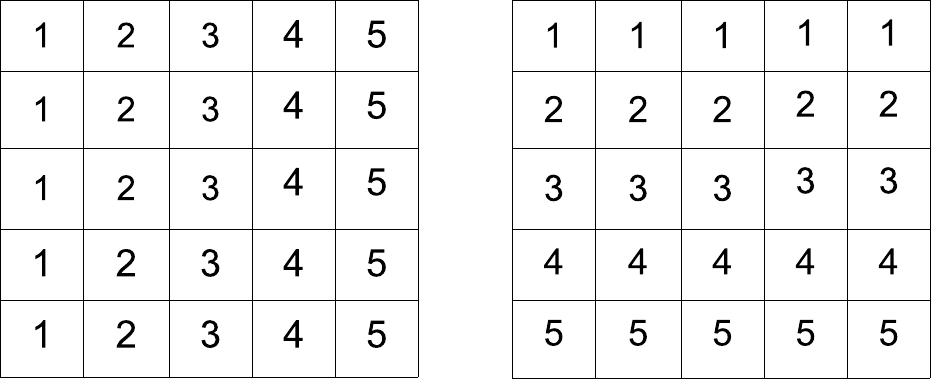}
\end{center}
\caption{Kernels used to compute the covariance array in $5 \times 5$ neighborhood: left) Kernel ($B^{i cent}$) to calculate $i$ element of the centroid; right) Kernel ($B^{j cent}$) to calculate $j$ element of the centroid.}
\label{Kernels}
\end{figure}

To extract the BEVM, principal components analysis (PCA) is used, which is applied on the local neighborhood of each power sample in the 2D representation. Hence, it aims at utilizing local description information in the 2D power matrix. Consequently, PCA \cite{Lever2017} is executed using square kernels. The covariance array of the local PCA can be generated as a group of second-order power matrix \cite{Hu1057692}. 
Power matrix moments are calculated just as the weighted average of power sample magnitudes of each power kernel, in which blocks of size $n \times n$ are considered. An example is portrayed in Fig. \ref{Kernels} representing kernels of size $5 \times 5$. 

Let $P_{i,j}$ be the 2D power matrix, the covariance array can be computed as follows:

\begin{equation}
\begin{array}{c}
Cov_{i,j}=\sum\limits_{u=1}^{u=n}\sum\limits_{v=1}^{v=n}\left(
B_{u,v}^{icent}-I_{i,j}^{cent}\right) \cdot \left(
B_{u,v}^{jcent}-J_{i,j}^{cent}\right) \cdot  \\ 
P_{_{i+u-\left( \frac{n-1}{2}\right) ,~j+v\left( \frac{n-1}{2}\right)
}}%
\end{array}%
\end{equation}
where $u$ and $v$ are the indexes used to handle the power samples in each kernel, $B_{u,v}^{i cent}$ and $B_{u,v}^{j cent}$ are the kernels used to calculate the $i$ element and $j$ element of the centroid, respectively, which are portrayed in Fig. \ref{Kernels}. $I_{i,j}^{cent}$ refers to the coordinates of the centroid calculated at each local neighborhood of a power sample at position $(i,j)$, in which blocks of size $n\times n$ samples are used.
\begin{equation}
I_{i,j}^{cent}=\frac{\sum\limits_{u=1}^{u=n}\sum\limits_{v=1}^{v=n}\left(
B_{u,v}^{icent}\cdot P_{_{i+u-\left( \frac{n-1}{2}\right) ,~j+v\left( 
\frac{n-1}{2}\right) }}\right) }{\sum\limits_{u=1}^{u=n}\sum%
\limits_{v=1}^{v=n}\left( P_{_{i+u-\left( \frac{n-1}{2}\right) ,~%
j+v\left( \frac{n-1}{2}\right) }}\right) }
\label{eq4}
\end{equation}
and $J_{i,j}^{cent}$ refers to the $J$ coordinates of the centroid calculated at each local neighborhood of a power sample at position $(i, j)$, in which blocks of size $n\times n$ samples are used.
\begin{equation}
J_{i,j}^{cent}=\frac{\sum\limits_{u=1}^{u=n}\sum\limits_{v=1}^{v=n}\left(
B_{u,v}^{jcent}\cdot P_{_{i+u-\left( \frac{n-1}{2}\right) ,~ j+v\left( 
\frac{n-1}{2}\right) }}\right) }{\sum\limits_{u=1}^{u=n}\sum%
\limits_{v=1}^{v=n}\left( P_{_{i+u-\left( \frac{n-1}{2}\right) ,~%
j+v\left( \frac{n-1}{2}\right) }}\right) }
\label{eq5}
\end{equation}

Therefore, the local covariance array is estimated as:
\begin{equation}
Cov\left( P_{i,j}\right) =\left[ 
\begin{array}{cc}
I_{i,j}^{var} & Cov_{i,j} \\ 
Cov_{i,j} & J_{i,j}^{var}%
\end{array}%
\right] 
\end{equation}
where $I_{i,j}^{var}$ and $J_{i,j}^{var}$ are the second-order moments implicated in the estimation of the covariance, they are defined as follows:
\begin{equation}
I_{i,j}^{var}=\sum \sum \left(
B_{u,v}^{i cent}-I_{i,j}^{cent}\right) ^{2}P_{i+u-\left( \frac{n-1}{2}%
\right) ,~j+v\left( \frac{n-1}{2}\right) }
\end{equation}
and 
\begin{equation}
J_{i,j}^{var}=\sum \sum \left(
B_{u,v}^{j cent}-J_{i,j}^{cent}\right) ^{2}P_{i+u-\left( \frac{n-1}{2}%
\right) ,~j+v\left( \frac{n-1}{2}\right) }
\end{equation}

Eigenvalue decomposition of the aforementioned covariance array results into two eigenvalues. Computing covariance array and decomposing it in terms of its eigenvalues are conducted on each power sample. Hence, an eigenvalue map is constructed with reference to the principal eigenvalues (PEV). 

It is worth noting that the size of the generated PEV matrix has been maintained the same as the 2D power matrix. Following, we normalize the PEV map via the calculation of the ratio of PEV to the power sample value, around which the local neighborhood has been investigated (center power sample). Moreover, the conducted normalization results in a constant PEV when it is calculated for a neighborhood of constant intensities. Specifically, for the regions of the power matrix with constant magnitudes, the normalized eigenvalue relies merely on the size of the neighborhood deployed to compute the local PCA. 

In this framework, kernels with a size of $15 \times 15$ (i.e. $n=15$) has been considered, where the normalized eigenvalue in the parts having a constant intensity is 4200. Furthermore, the normalization of PEV aids to set a threshold value for generating the BEVM. Using the normalized PEV for the parts having a constant intensity as the reference, the threshold is adjusted to be a bit higher to eliminate extremely fine details. Consequently, under this framework a threshold $thre = 4225$ is adopted empirically. The steps required for computing the BEVM are summarized in Algo. \ref{algo1}.

\begin{algorithm}[t!]
\SetAlgoLined
\KwResult{BEVM$_{i,j}$: binarized eigenvalue map}
1. Convert the power signal $p$ into a 2D representation to construct the power matrix $P_{i,j}$\\
 \While{$k \leq M$ (\textnormal{with} $M$ \textnormal{is the length of of the power matrix})}{

2. Calculate second-order moments of the power matrix ($I_{i,j}^{var}$ and $J_{i,j}^{var}$) using a block splitting process with kernels of size $n\times n$. \\

3. Estimate the local covariance array $Cov(P_{i_{k},j_{k}})$. \\

4. Apply an eigenvalue decomposition on the $Cov(P_{i_{k},j_{k}})$ to obtained PEV matrix. \\

5. Apply a normalization of the $PEV$ matrix obtained in Step 4 using the magnitude value of the centroid power sample ($P_{i_{k},j_{k}}$)

\begin{equation}
EVM_{i_{k},j_{k}} = Normalize(PEV)
\end{equation}
}
6. Apply a thresholding process on $EVM_{i,j}$ to generate $BEVM_{i,j}$ 
\begin{equation}
BEVM_{i,j}=\left\{ 
\begin{array}{cc}
1 & EVM_{i,j}\geq thre \\ 
0 & \textnormal{Else \ \ \ \ \ \ \ \ \ \ \ \ \ \ \ }%
\end{array}%
\right. 
\end{equation}

\caption{Proposed algorithm for computing BEVM.}
\label{algo1}
\end{algorithm}

Finally, the post-processing step is performed on the generated LBP image using the BEVM matrix extracted by Algo. \ref{algo1}. The objective is to extract two histograms called up and down histograms, in which $h_{up}=hist(LBP(BEVM=1)$ and $h_{down}=hist(LBP(BEVM=0)$. Following, the resulting histograms are then concatenated to build the overall histogram, $h_{LBP-BEVM}=[h_{up} ~ h_{down}]$, which represents the power signature of a specific appliance.


\subsection{Ensemble bagging tree (EBT) classifier}
\begin{figure}[t!]
\begin{center}
\includegraphics[width=8.1cm, height=8.6cm]{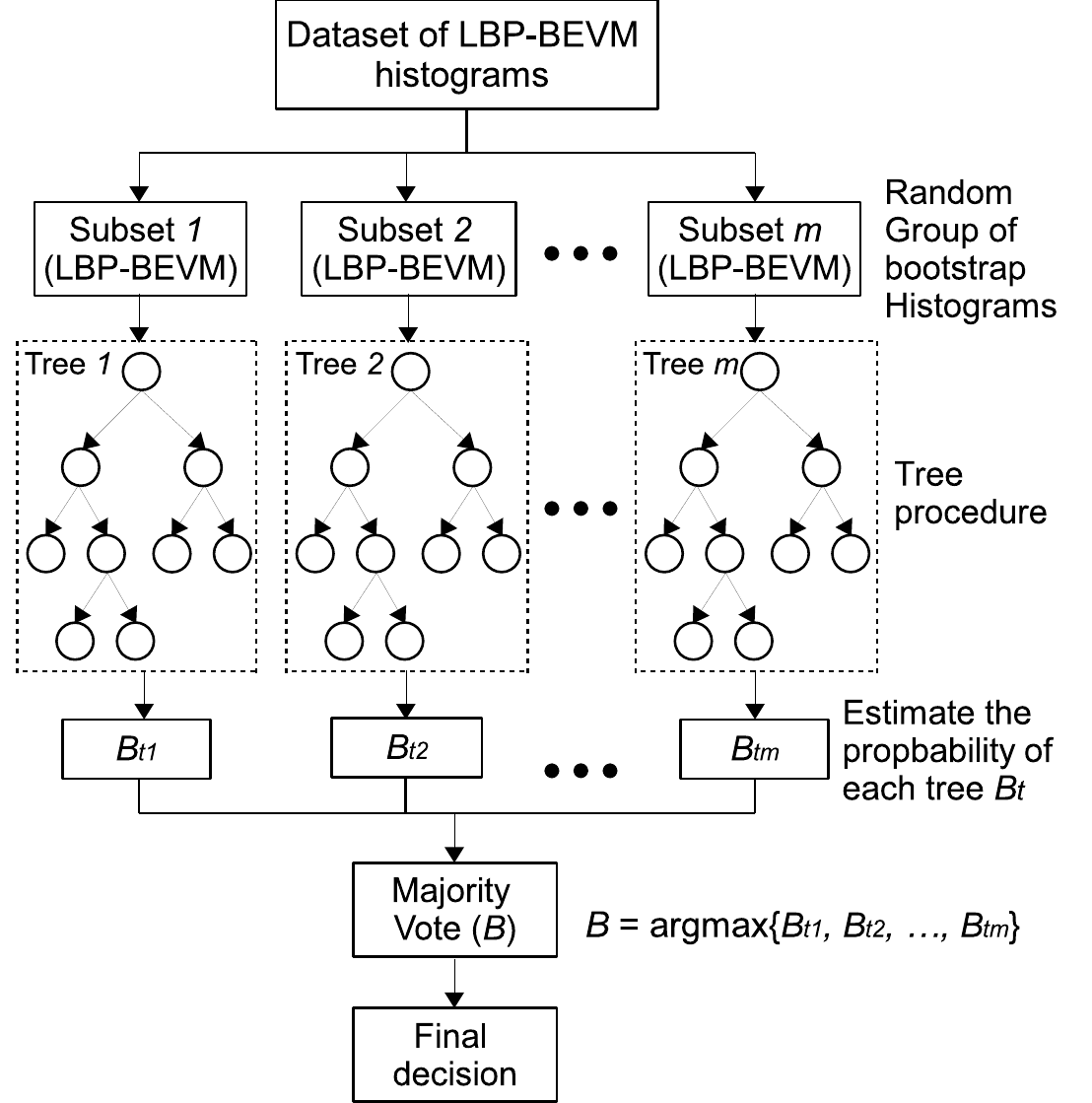}
\end{center}
\caption{Simple explanation of the EBT classifier}
\label{EBT}
\end{figure}

EBT is an efficient classifier that did not receive its merit in practice. Its importance comes from the fact that it can achieve a high
classification performance using a fusion of various weak classifiers. Therefore, LBP-BEVM histograms generated from power signals in a specific dataset are split into $m$ bootstrap subsets, in which each weak classifier is trained via a tree procedure, and hence a set of probabilities $B_{t1}, B_{t2}, \cdots, B_{tm}$ is generated. Following, a majority vote is performed to estimate the final probability $B$ as follows:
\begin{equation}
B = \arg \max (B_{t1}, B_{t2}, \cdots, B_{tm})
\end{equation} 
Afterward, the final classification decision is generated based on the estimated probability. Fig. \ref{EBT} portrays a simple explanation of the EBT classifier. 

\section{Experimental results} \label{sec4}
\subsection{Dataset description}
In order to assess the performance of the proposed appliance identification system based on LBP-BEVM, two datasets are used, defined as GREEND \cite{GREEND2014} and WHITED \cite{WHITED2016}. The first one includes electricity consumption signatures of several domestic appliances collected at a sampling rate of 1 Hz from 8 households in Italy and Austria. To validate the proposed system, power usage footprints gathered from a typical house are used for a period of 8 months. For WHITED dataset, we consider 11 appliance categories to validate the proposed system and each category includes various consumption fingerprints from distinct manufacturers, which are gathered at a sampling rate of 44000 Hz. Table \ref{WHITED} summarizes the electrical device categories considered to collect load consumption signatures for both WHITED and GREEND. 

\begin{table}[t!]
\caption{Description of monitored appliances on both the WHITED and GREEND datasets.}
\label{WHITED}
\begin{center}

\begin{tabular}{lll|lll}
\hline
\multicolumn{3}{c|}{WHITED} & \multicolumn{3}{|c}{GREEND} \\ \hline
Tag & device  & \# tested  & Tag & device & \# checked \\ 
& category & devices &  & Category & \multicolumn{1}{r}{days} \\ \hline
1 & Modem/receiver & \multicolumn{1}{r|}{20} & 1 & Coffee machine & 
\multicolumn{1}{r}{242} \\ 
2 & CFL & \multicolumn{1}{r|}{20} & 2 & Radio & \multicolumn{1}{r}{242} \\ 
3 & Charger & \multicolumn{1}{r|}{30} & 3 & Fridge / freezer & 
\multicolumn{1}{r}{240} \\ 
4 & Coffee machine & \multicolumn{1}{r|}{20} & 4 & Dishwasher & 
\multicolumn{1}{r}{242} \\ 
5 & Drilling machine & \multicolumn{1}{r|}{20} & 5 & kitchen lamp & 
\multicolumn{1}{r}{242} \\ 
6 & Fan & \multicolumn{1}{r|}{30} & 6 & TV & \multicolumn{1}{r}{242} \\ 
7 & Flatron & \multicolumn{1}{r|}{20} &  &  &  \\ 
8 & LED ight & \multicolumn{1}{r|}{20} &  &  &  \\ 
9 & Kettle & \multicolumn{1}{r|}{20} &  &  &  \\ 
10 & Microwave & \multicolumn{1}{r|}{20} &  &  &  \\ 
11 & Iron & \multicolumn{1}{r|}{20} &  &  &  \\ \hline
\end{tabular}

\end{center}
\end{table}

\subsection{Parameter setting}
\begin{figure}[t!]
\begin{center}
\includegraphics[width=8.9cm, height=3.6cm]{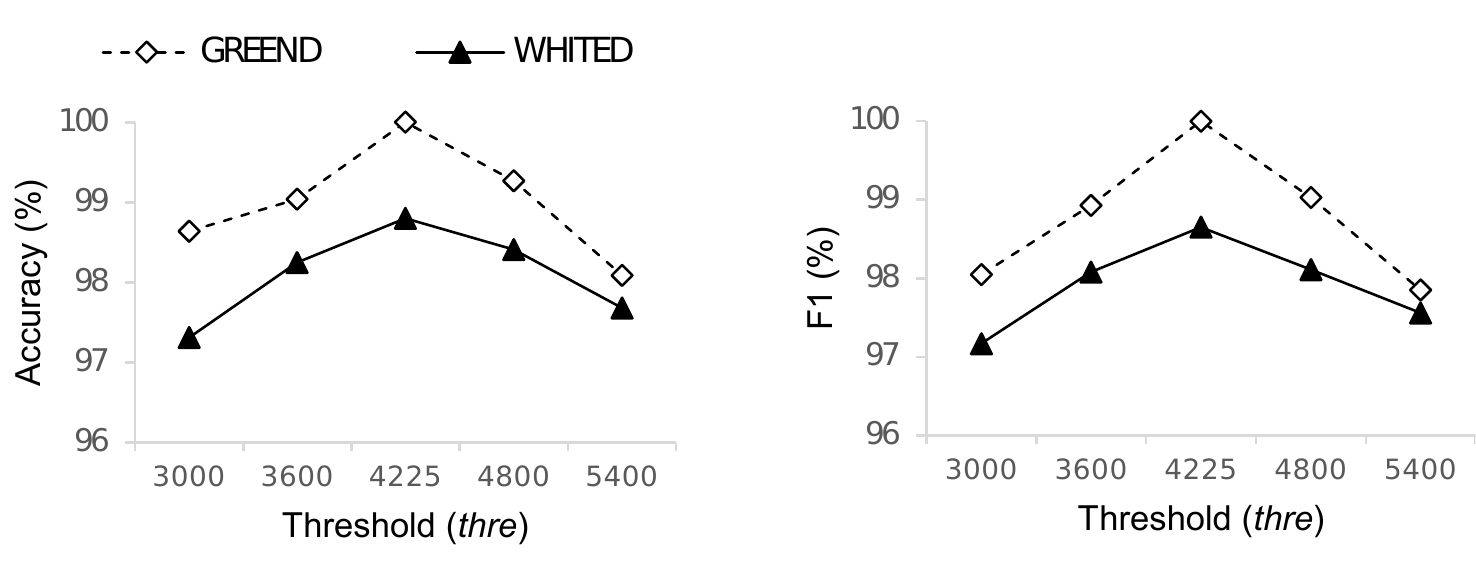}
\caption{Impact of varying the threshold ($thre$) on appliance identification performance: left) accuracy and right) F1 score.}
\label{ConstMagnitude}
\end{center}
\end{figure}

\begin{figure}[t!]
\begin{center}
\includegraphics[width=8.9cm, height=3.6cm]{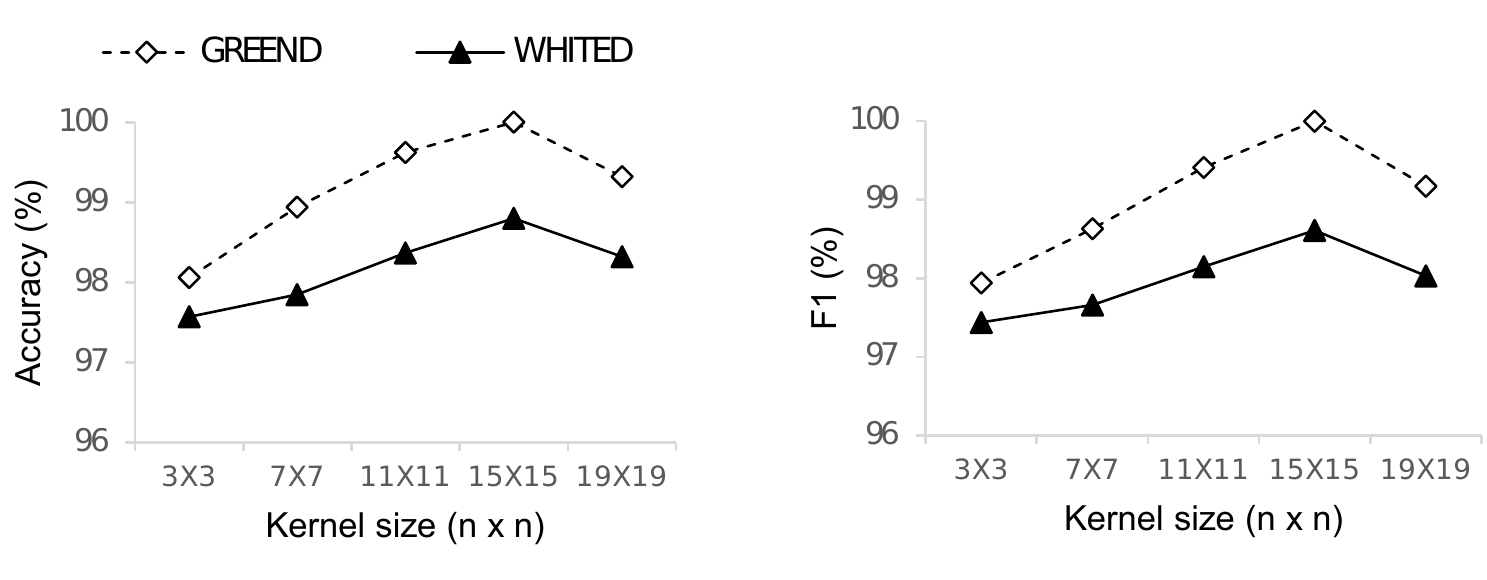}
\caption{Impact of varying the kernel size on appliance identification performance: left) accuracy and right) F1 score.}
\label{KernelSize}
\end{center}
\end{figure}

In this framework, a kernel of $15 \times 15$ (i.e. $n=15$) with a $thre = 4225$ are selected empirically. In order to justify this selection, their values have been changed empirically in order to check what are the optimized values. Fig. \ref{ConstMagnitude} illustrates the effect of varying the values of the  threshold $thre$ on the performance of appliance identification in terms of the accuracy and F1 score. While Fig. \ref{KernelSize} portrays the impact of varying the values of the kernel size $n \times n$ on the classification performance in terms of the accuracy and F1 score as well. In this context, it can be seen that for the case of the threshold, the value $thre = 4225$ provides the best accuracy and F1 score performance. On the other side, the kernel size of $15 \times 15$ allows to obtain the highest accuracy and F1 score rates. 

\subsection{Comparison of machine learning classifiers}
We conduct a performance comparison of EBT classifier versus other machine learning models, including support vector machines (SVM), deep neural networks (DNN), K-nearest neighbors (KNN), decision tree (DT) operating by reference to different classification parameters, implemented using Matlab 2018a and running on a desktop having a Core i7-3770S processor, 16 GRAM and 3.1 GHz. 
Table \ref{classComp} illustrates the accuracy and F1 score outputs obtained for both the GREEND and WHITED datasets where a threshold $thre = 4225$ is adopted. It is clearly shown that the EBT outperformed the other classifiers with respect to the accuracy and F1 score.

\begin{table}[t!]
\caption{Accuracy and F-score of the EBT model compared with other classifiers using LBP-BEVM.}
\label{classComp}
\begin{center}

\begin{tabular}{lccccc}
\hline
\textbf{ML } & \textbf{Classifier} & \multicolumn{2}{c}{\textbf{GREEND}} & 
\multicolumn{2}{c}{\textbf{WHITED}} \\ \cline{3-6}\cline{3-5}
\textbf{classifier} & \textbf{\ parameters} & \textbf{acc} & \textbf{F1}
& \textbf{acc} & \textbf{F1} \\ \hline
{\small LDA} &  & 96.85 & 96.42 & 96.75 & 96.28 \\ 
{\small SVM} & {\small Linear Kernel} & 88.78 & 88.56 & 90.83 & 89.01 \\ 
{\small SVM} & {\small Quadratic kernel} & 84.69 & 83.54 & 95.41 & 95.02 \\ 
{\small SVM} & {\small Gaussian kernel} & 89.75 & 89.32 & 93.33 & 92.92 \\ 
{\small KNN} & {\small K=1, Euclidean dist} & 95.68 & 95.41 & 96.25 & 94.95
\\ 
{\small KNN} & {\small K=10, Weighted} & 94.83 & 94.58 & 95.83 & 94.55 \\ 
& {\small Euclidean dist} &  &  &  &  \\ 
{\small KNN} & {\small K=10, Cosine dist} & 92.46 & 91.1 & 89.16 & 87.11 \\ 
{\small DT} & {\small Fine, 100 splits} & 93.8 & 93.69 & 95 & 94.01 \\ 
{\small DT} & {\small Medium, 20 splits} & 92.41 & 92.17 & 94.16 & 92.02 \\ 
{\small DT} & {\small Coarse, 4 splits} & 67.72 & 63.39 & 35.41 & 29.35 \\ 
{\small DNN} & {\small 50 hidden layers} & 96.22 & 96.14 & 95.37 & 94.89 \\ 
\textbf{EBT} & \textbf{30 learners, 42 k} & \textbf{100} & \textbf{100} & 
\multicolumn{1}{l}{\textbf{98.8}} & \multicolumn{1}{l}{\textbf{98.65}} \\ 
& \textbf{\ splits} &  &  &  &  \\ \hline
\end{tabular}

\end{center}
\end{table}

\subsection{Comparison with other 2D descriptors}
\begin{figure}[t!]
\begin{center}
\includegraphics[width=8cm, height=3.4cm]{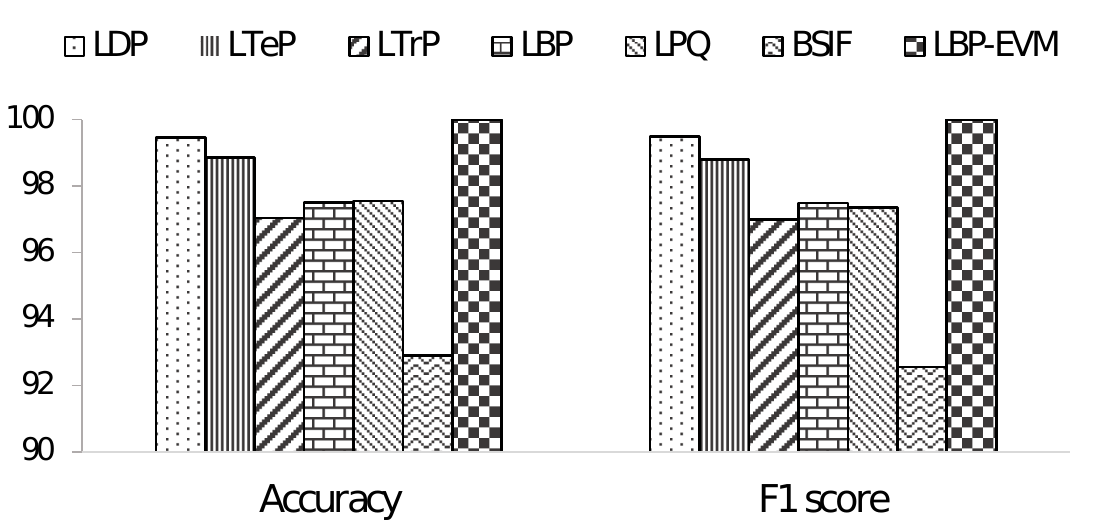}
\includegraphics[width=8cm, height=3cm]{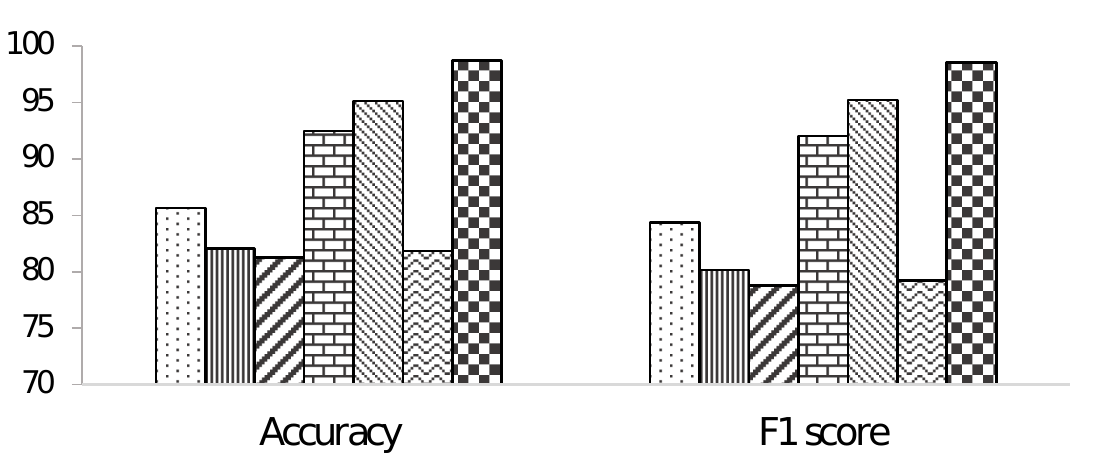}

\end{center}
\caption{Accuracy and F1 score comparison of LBP-BEVM descriptor with other 2D descriptors under the: a) GRENND and b) WHITED datasets.}
\label{Results}
\end{figure}

To highlight the improvement introduced due the post-processing stage in LBP-BEVM, performance is compared with other well-known 2D descriptors that are widely used in image processing, including local directional patterns (LDP) \cite{Perumal2016}, local ternary pattern (LTeP) \cite{Yuan2014}, local transitional pattern (LTrP) \cite{Ahsan2013}, local phase quantization (LPQ) \cite{RAHTU2012501}, binarized statistical image features (BSIF) \cite{Kannala6460393} and LBP. Table \ref{LBPvriants} and Fig. \ref{Results} depict the comparison results of the aforementioned descriptors under GREEND and WHITED datasets in terms of the histogram length, accuracy and F1 score with reference to EBT classifier. 
It is clearly witnessed that LBP-BEVM outperforms the other description schemes for both the accuracy and F1 score. Further, LBP-BEVM histogram that includes 112 samples, has a much lower length compared to the other descriptors (LTeP, LTrP, LBP, LPQ and BSIF) except for the case of the LDP, which has a length of 56 bins.  

It can also be deduced that all descriptors except the BSIF perform well on GREEND dataset, they have accuracies and F1 scores of more than 96\%. In which, LBP-BEVM reaches 100\% for both the accuracy and F1 score. However, their performance are dramatically dropped under WHITED dataset and only LBP-BEVM keeps a high efficiency. Specifically, up to 98.8\% accuracy and 98.65\% F1 score are achieved. In addition, we notice that the superiority of LBP-BEVM in comparison to LBP, LPQ, LTrP, LTeP, LDP and BSIF descriptors can reach more than 3\%, 6\%, 17\%, 16\%, 13\% and 16\%, respectively, in terms of the accuracy and F1 score.

\begin{table}[t!]
\caption{Comparison of LBP-BEVM descriptor with other 2D descriptors in terms of the histogram length, accuracy and F1 score. }
\label{LBPvriants}
\begin{center}

\begin{tabular}{ccllll}
\hline
{\scriptsize \textbf{Algorithm}} & {\scriptsize \textbf{Histogram}} & \multicolumn{2}{c}{\scriptsize \textbf{GREEND}}
& \multicolumn{2}{c}{\scriptsize \textbf{WHITED}} \\ \cline{3-6}
\multicolumn{1}{l}{} & \multicolumn{1}{l}{\scriptsize \textbf{length}} & {\scriptsize \textbf{accuracy}}
& {\scriptsize \textbf{F1 score}} & {\scriptsize \textbf{accuracy}} & {\scriptsize \textbf{F1 score}} \\ \hline
\multicolumn{1}{l}{\scriptsize LDP} & \multicolumn{1}{l}{\scriptsize 56} & {\scriptsize %
99.46} & {\scriptsize 99.50} & {\scriptsize 85.66} & {\scriptsize 84.38} \\ 
\multicolumn{1}{l}{\scriptsize LTeP} & \multicolumn{1}{l}{\scriptsize 512} & {\scriptsize %
98.86} & {\scriptsize 98.80} & {\scriptsize 82.08} & {\scriptsize 80.15} \\ 
\multicolumn{1}{l}{\scriptsize LTrP} & \multicolumn{1}{l}{\scriptsize 256} & {\scriptsize %
97.04} & {\scriptsize 96.99} & {\scriptsize 81.25} & {\scriptsize 78.78} \\ 
\multicolumn{1}{l}{\scriptsize LBP} & \multicolumn{1}{l}{\scriptsize 256} & {\scriptsize %
97.50} & {\scriptsize 97.49} & {\scriptsize 92.5} & {\scriptsize 92.04} \\ 
\multicolumn{1}{l}{\scriptsize LPQ} & \multicolumn{1}{l}{\scriptsize 256} & {\scriptsize %
97.56} & {\scriptsize 97.35} & {\scriptsize 95.16} & {\scriptsize 95.24} \\ 
\multicolumn{1}{l}{\scriptsize BSIF} & \multicolumn{1}{l}{\scriptsize 256} & {\scriptsize %
92.91} & {\scriptsize 92.55} & {\scriptsize 81.85} & {\scriptsize 79.21} \\ 
\multicolumn{1}{l}{\scriptsize \textbf{LBP-EVM}} & \multicolumn{1}{l}{\scriptsize 112} & 
{\scriptsize \textbf{100}} & {\scriptsize \textbf{100}} & {\scriptsize \textbf{98.8}} & {\scriptsize \textbf{98.65}} \\ \hline
\end{tabular}

\end{center}
\end{table}

\subsection{Correlation and discrimination ability}
\begin{figure}[t!]
\begin{center}
\includegraphics[width=8.8cm, height=4.1cm]{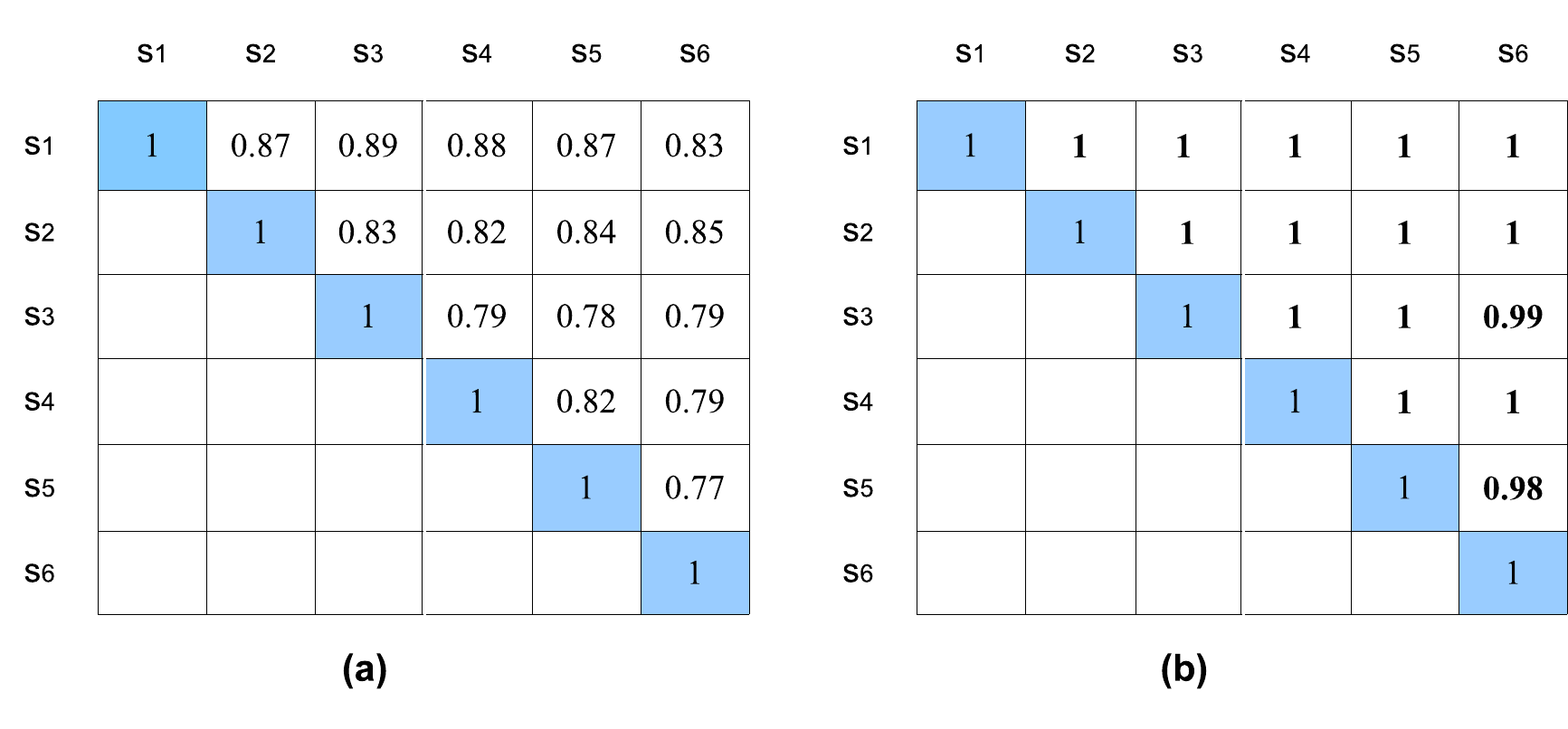}
(I) Kettle\\
\includegraphics[width=8.8cm, height=4.1cm]{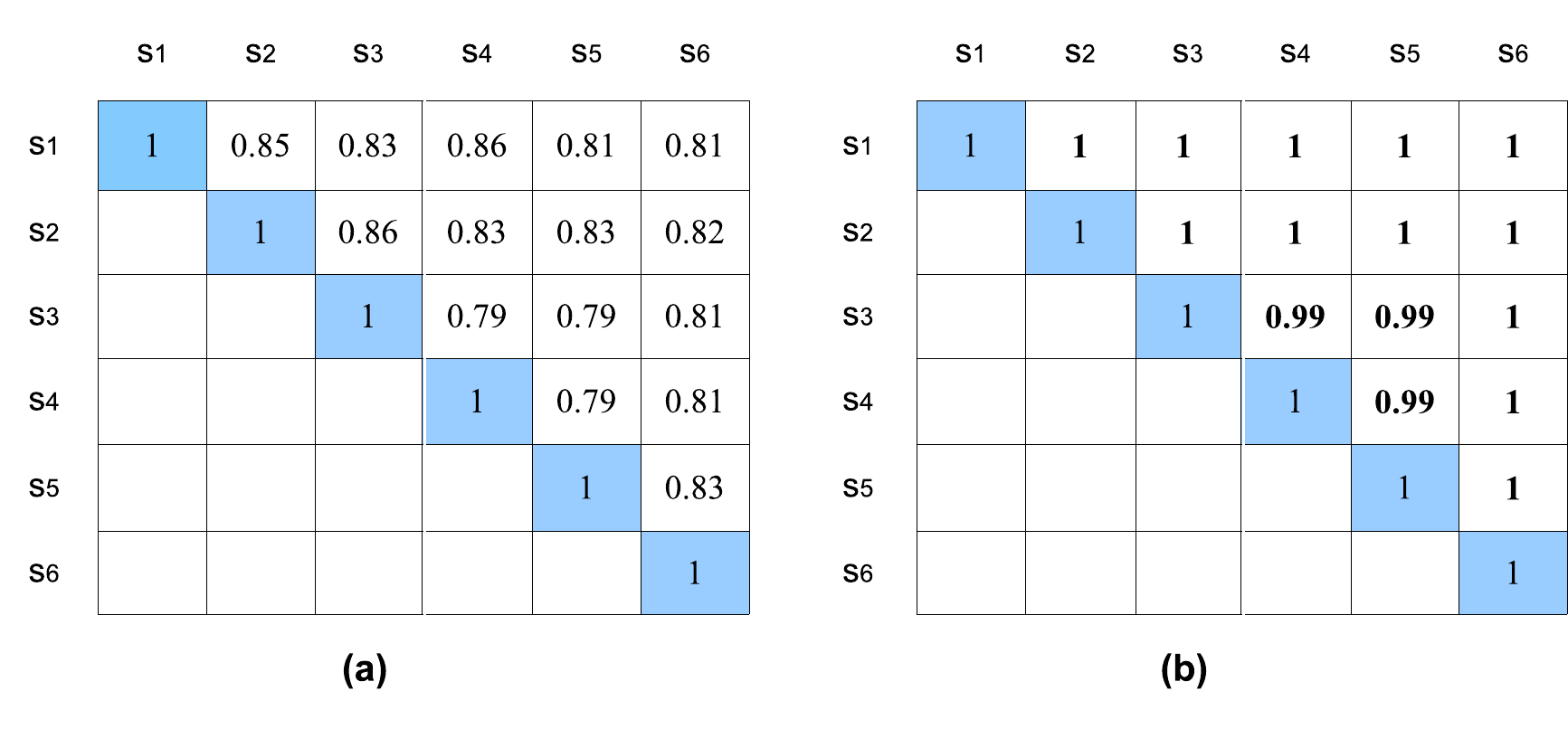}
(II) Washing machine\\
\end{center}
\caption{NCC performance between the appliances from the same classes of (a) LBP descriptor and (b) LBP-BEVM.}
\label{CorrMat}
\end{figure}

It is of paramount importance to comprehend why the proposed LBP-BEVM descriptor performs well for appliance identification and outperforms the other descriptors, especially the conventional LBP. Therefore, this section investigates how BEVM can help LBP-BEVM to correlate efficiently between appliances pertaining to the same class, and on the other side to what extent it can discriminate between appliances belonging to different classes. First, two appliance classes are selected from WHITED dataset, in which six signals are then designated randomly from each appliance class $s_{1}, s_{2}, \cdots , s_{6}$. Next, the normalized cross-correlation (NCC) rates between the LBP features extracted from those signals and on the flipside between LBP-BEVM features are calculated, respectively. If we consider $F^{1}$ and $F^{2}$ to be the feature vectors derived from $s_{1}$ and $s_{2}$, respectively, the NCC between these two vectors is given as follows:
\begin{equation}
NCC=\frac{F^{1}\cdot F^{2}}{\left\vert F^{1}\right\vert \left\vert
F^{2}\right\vert }=\frac{\sum\nolimits_{i}F^{1}_{i}\cdot F^{2}_{i}}{\sqrt{%
\sum\nolimits_{i}F^{1}_{i}}\sqrt{\sum\nolimits_{i}F^{2}_{i}}}, ~~-1\leq NCC\leq 1
\end{equation}
Fig. \ref{CorrMat} portrays NCC matrices computed using (a) LBP features, and (b) LBP-BEVM descriptions extracted from six signals pertaining to the coffee machine and washing machine classes, respectively. It is clearly seen that NCC values between LBP features are quietly low and can not be greater than 0.89, while those obtained with LBP-BEVM descriptions can reach 1 between most of the signals. Consequently, it can be deduced that LBP-BEVM correlates effectively and better than LBP between appliances pertaining to the same category.

\begin{figure}[t!]
\begin{center}
\includegraphics[width=8.8cm, height=4.1cm]{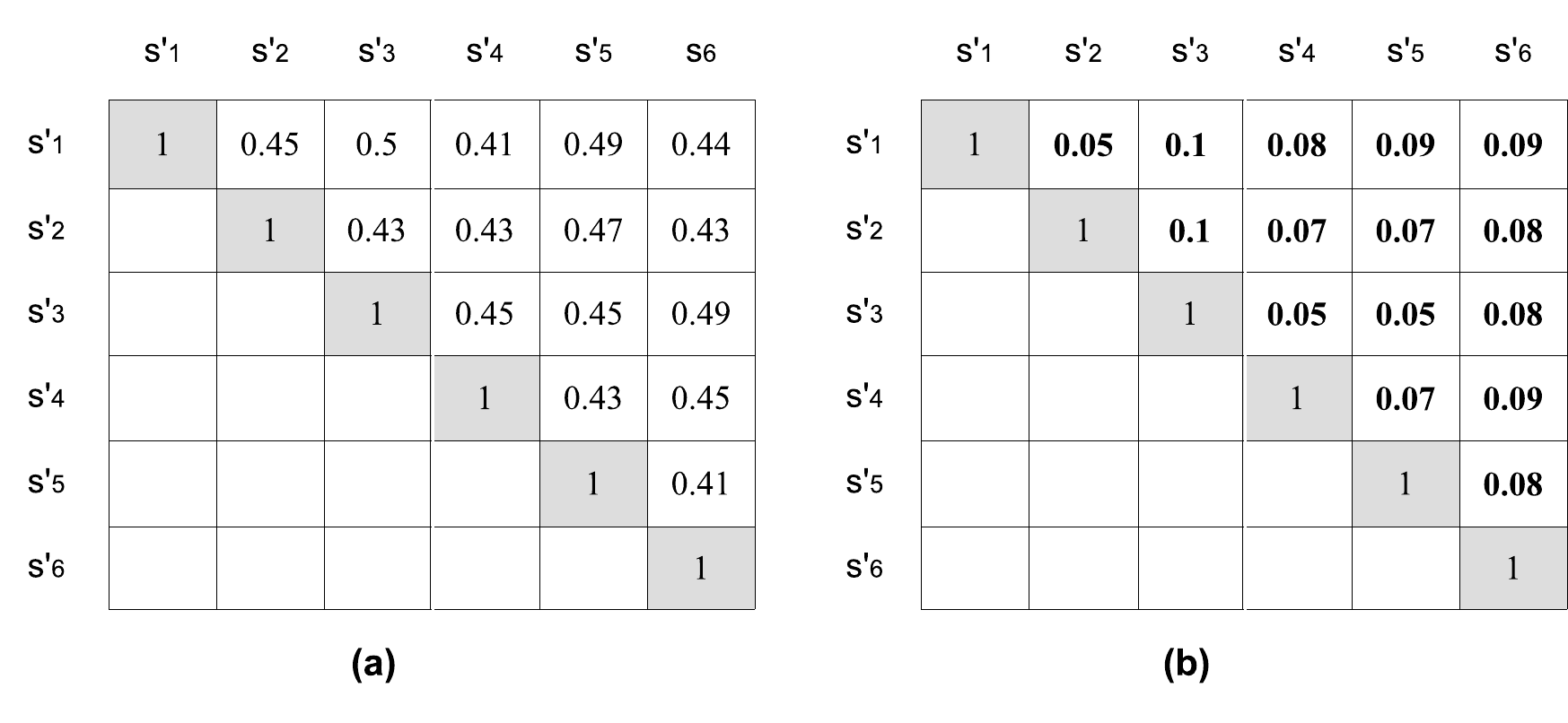}
\end{center}
\caption{NCC performance between appliances from distinct classes of (a) LBP descriptor and (b) LBP-BEVM.}
\label{DiscMat}
\end{figure}

On the other side, we select six different signals $s'_{1}, s'_{2}, \cdots , s'_{6}$ from distinct appliance classes collected in WHITED dataset, which are defined as the coffee machine, fan, LED light, kettle, microwave, and iron. Moving forward, the NCC values are estimated between them to check the discrimination ability of BEVM-LBP compared to conventional LBP. Fig. \ref{DiscMat} portrays the NCC matrices computed using (a) LBP features, and (b) BP-BEVM descriptions, respectively. It is clearly shown that LBP-BEVM has a better discrimination ability since it has the lowest correlation rates between the different signals from distinct classes. Explicitly, NCC values are lower than 0.1, while for conventional LBP NCC values are more than 0.4. 
Both case study scenarios addressed in this section prove the aptitude of LBP-BEVM for encoding effectively the appliance class-specific features by removing the unnecessary information, and therefore leading to a better correlation of appliance features from the same class, while resulting in a very low correlation between appliances fingerprints from distinct classes.

\begin{figure}[t!]
\begin{center}
\includegraphics[width=5.4cm, height=3.6cm]{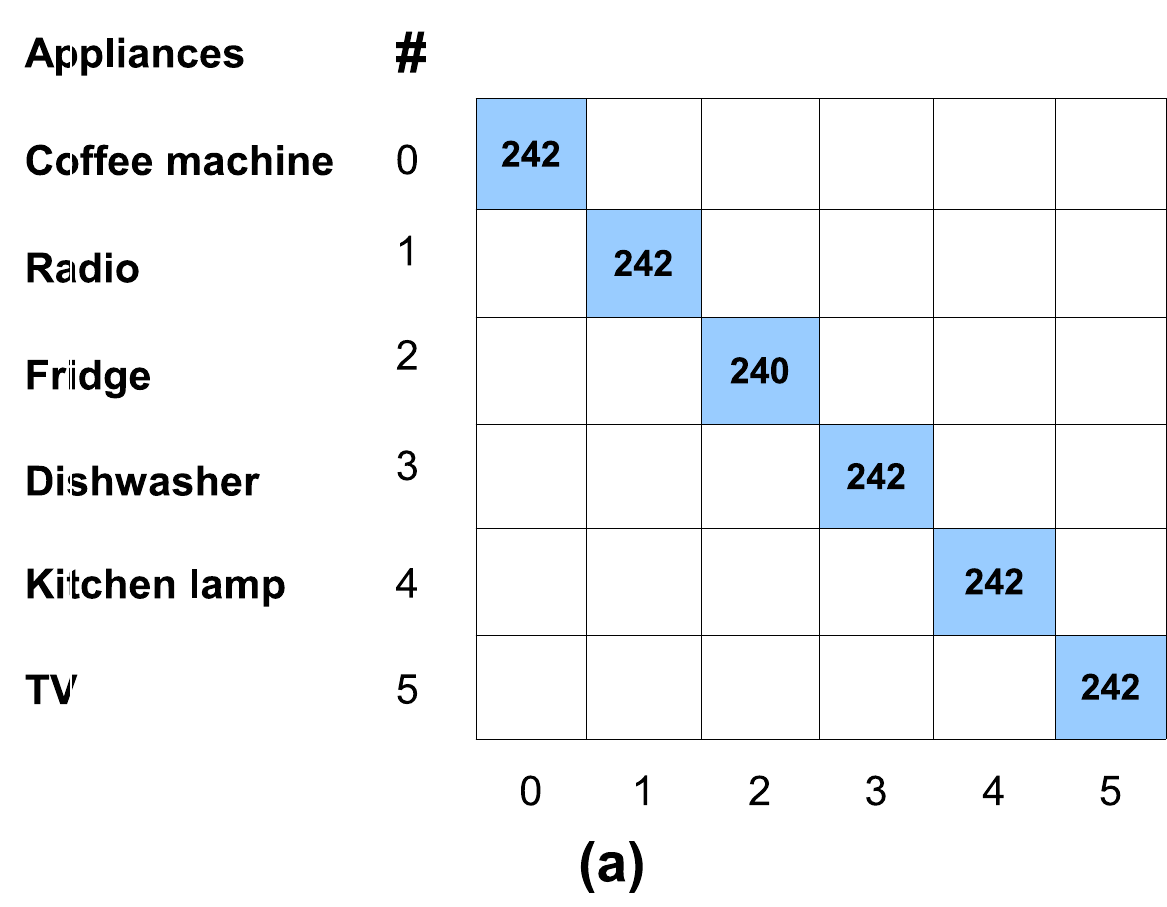}
\includegraphics[width=8.5cm, height=6.8cm]{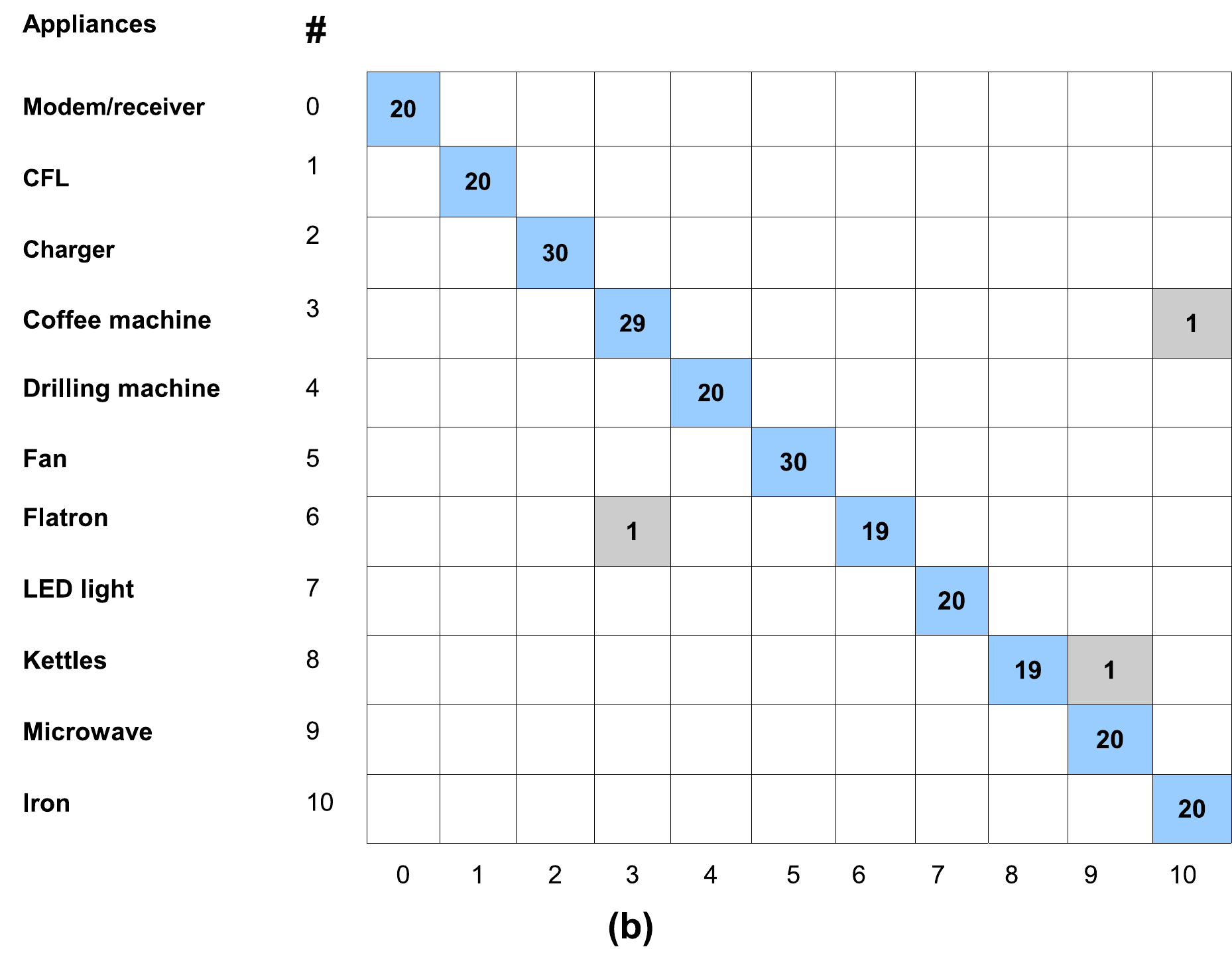}
\end{center}
\caption{Confusion matrix of LBP-BEVM descriptor using DBT classifier for: (top) GREEND and (down) WHITED.}
\label{ConMat}
\end{figure}

Confusion matrices of the LBP-BEVM based system obtained under GREEND and WHITED datasets are drawn in Fig. \ref{ConMat}. These results are achieved while the EBT classifier has been adopted. It is clearly seen from Fig. \ref{ConMat}(a) that the proposed solution can perfectly recognize all the appliances considered under GREEND dataset using daily signatures collected through a specific period (i.e. nearly 8 months). Moreover, high performance are also obtained under WHITED dataset, in which only 3 from 250 appliances are misclassified, which are mainly pertaining to the coffee machine, Flatron and kettle appliance classes, as it is portrayed in Fig. \ref{ConMat}(b).     

\subsection{Comparison with existing appliance identification systems}
To further assess the performance of the proposed LBP-BEVM based appliance identification system with other state-of-the-art solutions, Table \ref{AppIdenComp} summarizes another comparison in terms of the adopted technique, accuracy and F1 score. 
It can be observed that the proposed system presents the best performance, in which more than 2\% accuracy and F1-score improvements have been achieved in comparison the MSWPT based solution \cite{Himeur2020AE}. Further, more than 9\%, 5\% and 4\% accuracy and F1 score improvements have been attained compared to the frameworks described in \cite{Park8031399}, \cite{Le9042316} and \cite{Xiao8873557}.

\subsection{Computational Cost}
The computational cost of LBP-BEVM based appliance identification solution has been evaluated versus the other recent frameworks. All algorithms considered in this comparison are implemented using Python 3.7, they have been running on a desktop, having a Core i7-3770S processor, 16 GRAM and 3.1 GHz. Table \ref{TimeCompl} depicts the obtained results, where it is clearly seen that conventional LBP, described in \cite{Ahonen2006}, has the lowest computation cost and BEVM slightly increases the computation time of LBP-BEVM compared to LBP. This proves that the complexity introduced by BEVM is negligible if we consider their accuracy and F1 score improvements. Moreover, the low test time attained with LBP-BEVM demonstrates that this approach is a potential candidate for real-time applications, since only 0.07 sec is required to identify a query appliance.

\section{Conclusion} \label{sec5}
An original scheme based on transforming power signals into 2D space and applying an improved LBP descriptor has been proposed in this paper. While extracting 2D LBP representation, another segment of information has been generated in parallel using BEVM, which is then deployed to extract two new histograms. BEVM relies on analyzing the neighbors of each center pixel using a specific patch with the aim of generating a binary information that has been used to improve the LBP histogram generation. The obtained histograms have then been concatenated to form the overall histogram. The superiority of LBP-BEVM versus other 2D descriptors and appliance identification systems has been demonstrated through a series of tests. Our future work will aim at implementing LBP-BEVM in a real-time NILM application, in which appliance-data can be retrieved from aggregated records.

\begin{table}[t!]
\caption{Comparison with existing appliance identification systems.}
\label{AppIdenComp}
\begin{center}

\begin{tabular}{cccc}
\hline
\textbf{Work} & \textbf{Feature} & \textbf{Accuracy} & \textbf{F1 score} \\ 
&  & \textbf{(\%)} & \textbf{(\%)} \\ \hline
\multicolumn{1}{l}{\cite{Park8031399}} & real power features + CCNN & 89.79
& 88.47 \\ 
\multicolumn{1}{l}{\cite{Du6982196}} & finite-state-machine characteristics
& 95.3 & - \\ 
\multicolumn{1}{l}{\cite{Xiao8873557}} & statistical model + KM agorithm & 
94.22 & 94.05 \\ 
\multicolumn{1}{l}{\cite{Le9042316}} & lower current harmonic + FFT & 93.81
& 93.27 \\ 
\multicolumn{1}{l}{\cite{Himeur2020AE}} & MSWPT & 96.47 & 96.31 \\ 
\multicolumn{1}{l}{\textbf{Proposed }} & LBP-BEVM & \textbf{98.8} & \textbf{%
98.65} \\ \hline
\end{tabular}

\end{center}
\end{table}

\begin{table} [t!]
\caption{Time complexity of the proposed solution compared to other recent frameworks.}
\label{TimeCompl}
\begin{center}

\begin{tabular}{lll}
\hline
\textbf{Work} & \textbf{Training time (in sec)} & \textbf{Test time (in sec)}
\\ \hline
\cite{Ahonen2006} & \multicolumn{1}{c}{\textbf{7.21}} & \multicolumn{1}{c}{%
\textbf{0.05}} \\ 
\cite{Park8031399} & \multicolumn{1}{c}{39.11} & \multicolumn{1}{c}{1.03} \\ 
\cite{Du6982196} & \multicolumn{1}{c}{31.79} & \multicolumn{1}{c}{0.76} \\ 
\cite{Xiao8873557} & \multicolumn{1}{c}{13.5} & \multicolumn{1}{c}{0.14} \\ 
\cite{Himeur2020AE} & \multicolumn{1}{c}{23.2} & \multicolumn{1}{c}{0.23} \\ 
\textbf{Proposed} & \multicolumn{1}{c}{9.76} & \multicolumn{1}{c}{0.07} \\ 
\hline
\end{tabular}

\end{center}
\end{table}

\section*{Acknowledgements}\label{acknowledgements}
This paper was made possible by National Priorities Research Program (NPRP) grant No. 10-0130-170288 from the Qatar National Research Fund (a member of Qatar Foundation). The statements made herein are solely the responsibility of the authors.


\end{document}